\documentclass[aps,10pt,superscriptaddress,pre,twocolumn,showpacs]{revtex4}
\usepackage{graphicx,epstopdf,epsfig,graphics,amssymb,amsmath,subeqnarray,color}
\begin{document}

\title{Revisiting the Saffman-Taylor experiment:  imbibition patterns and liquid-entrainment transitions}
\author{Bertrand Levach\'e}
\affiliation{PMMH-CNRS--ESPCI ParisTech-Universit\'e Paris 6-Universit\'e Paris 7, 10, rue Vauquelin, 75005 Paris, France}
\affiliation{Total  France, Pole d'Etudes et de Recherche de Lacq, BP47-64170, Lacq, France}
\author{Denis Bartolo}
\affiliation{PMMH-CNRS--ESPCI ParisTech-Universit\'e Paris 6-Universit\'e Paris 7, 10, rue Vauquelin, 75005 Paris, France}
\affiliation{Ecole Normale Sup\'erieure de Lyon, CNRS, 46, all\'e d'Italie, 69007 Lyon, France}
\begin{abstract}
We revisit  the Saffman-Taylor experiment focusing on the forced-imbibition regime where the displacing fluid wets the confining walls.  We demonstrate a new class of invasion patterns that do not display the canonical fingering shapes. We evidence that these unanticipated patterns stem from  the entrainement of thin liquid films from the moving meniscus. We then theoretically explain how  the interplay between the fluid flow at the contact line and the interface deformations  results  in the destabilization of liquid interfaces moving past solid surfaces. In addition, this minimal model conveys a unified framework  which consistently accounts for all the liquid-entrainment scenarios that have been  hitherto reported.
\end{abstract}
\maketitle

What liquid  should be used to clean a hydrophilic container filled with an organic fluid?
This seemingly trivial question turns out to be of major importance in a number of industrial process, including enhance  recovery of the so-called  heavy oils. An elementary thermodynamic reasoning  would suggest using an aqueous liquid making the smallest possible contact angle with the container walls. In this letter we show that the answer is actually  more subtle when the dynamics of  the fluid interfaces is considered. 

From a fundamental perspective,  liquid-liquid interfaces  driven past solid substrates have been extensively used as  a proxy to investigate nonlinear-pattern formation such as Laplacian growth processes~\cite{batchelor2002perspectives,Bensimon1986,HOMSY:2002wb,Lenormand:2001wh}. Until now  the overwhelming majority of the experiments have been performed in the drainage regime, where a low-viscosity fluid displaces a high-viscosity fluid which preferentially wets the solid. From    the Saffman-Taylor fingers growing in Hele-Shaw cells~\cite{Saffman1958,batchelor2002perspectives,Bensimon1986} to the fractal  patterns found in porous media~\cite{HOMSY:2002wb,SAHIMI:1993ta,Lenormand:2001wh}, the  salient features of all the drainage patterns are very well  captured by coarse-grained front-propagation models that discard the very details of the interactions between the liquid and the solid walls. 
Conversely, the experiments on  imbibition dynamics, where the less viscous phase preferentially wets the solid walls, have been  scarce and have yield somehow puzzling results
~\cite{Weitz:1987vz,Stokes:1986wc,Aarts2013}. The first quantitative experiment in a prototypal Hele-Shaw geometry was performed only   one year ago with colloidal liquids~\cite{Aarts2013}. Confocal imaging revealed an instability of the contact line. However, the resulting  entrainement of a thin liquid sheet  does not qualitatively modify the shape of viscous fingers. In contrast, imbibition experiments in porous media had revealed a marked qualitative change in the morphologies of the invasion patterns~\cite{Weitz:1987vz,Stokes:1986wc,Lenormand:2001wh}
\begin{figure}
	\includegraphics[width=\columnwidth]{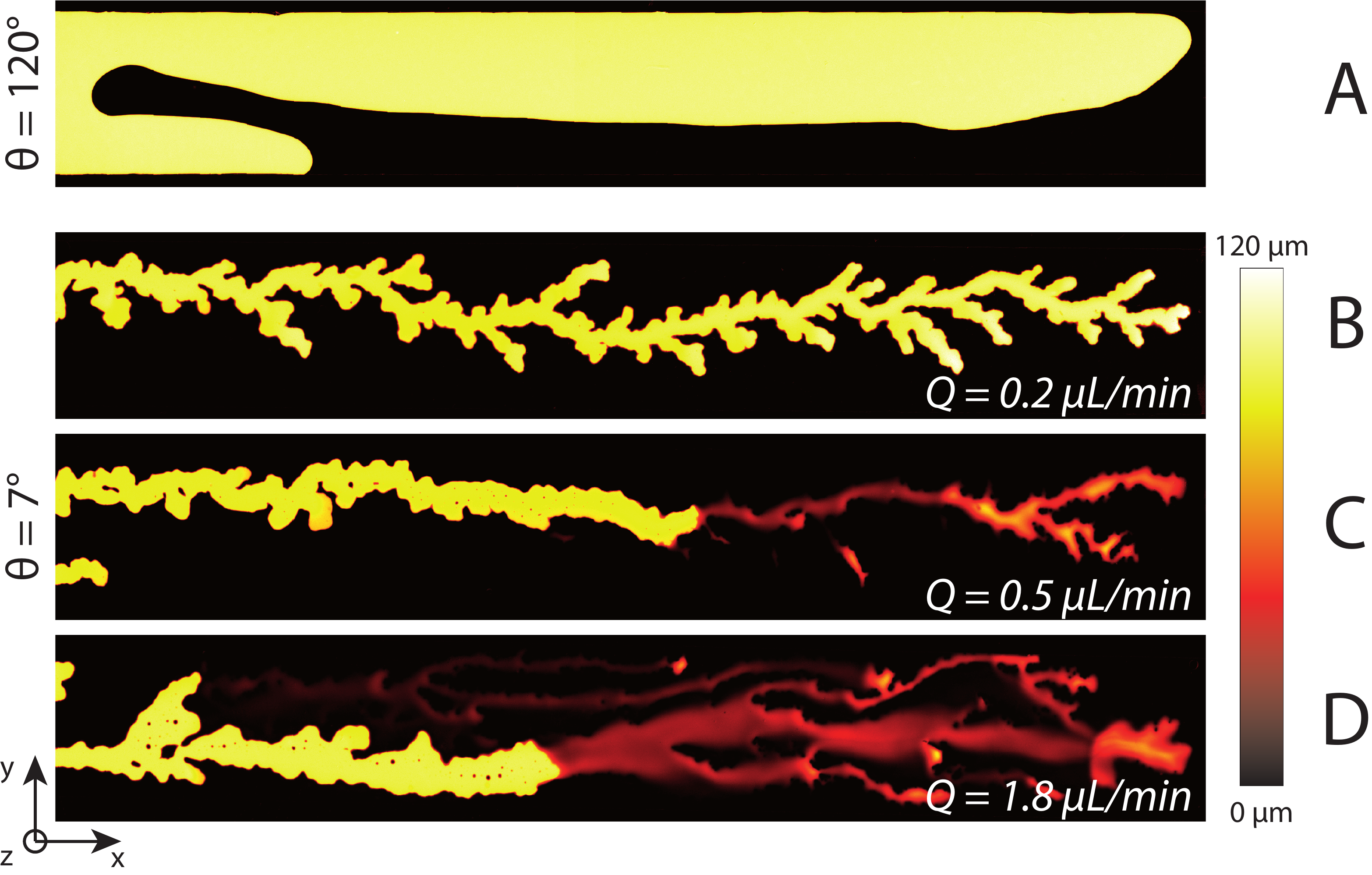}
	\caption{Viscous fingering pattern, and associated water-thickness fields, $\eta_{\rm oil}= 10^{3}\,{\rm cp}$. A- Drainage pattern, $Q=0.5\,\mu{\rm l/s}$. B, C and D-Imbibition patterns observed at three different flow rates.  See also the corresponding supplementary movies 1, 2 and 3.}
\label{fig1}
\end{figure}

Here, we revise the seminal Saffman-Taylor experiment using water to mobilize  viscous oils filling  hydrophilic  microfluidic channels. We demonstrate a novel type of liquid-entrainement instability and the subsequent  growth  of  unanticipated imbibition patterns. We first quantitatively characterize their shape and propagation dynamics.   We then theoretically explain how  the intimate coupling between the short-scale molecular interactions with the solid and the large-scale flows  results  in the destabilization of the two-fluid interface.
This model conveys a unified framework  to consistently accounts for all the liquid entrainement scenarios that have been  reported so far~\cite{Snoeijer:2013cm,Aarts2013,Ledesma-Aguilar2013}.
 \begin{figure*}
	\includegraphics[width=\textwidth]{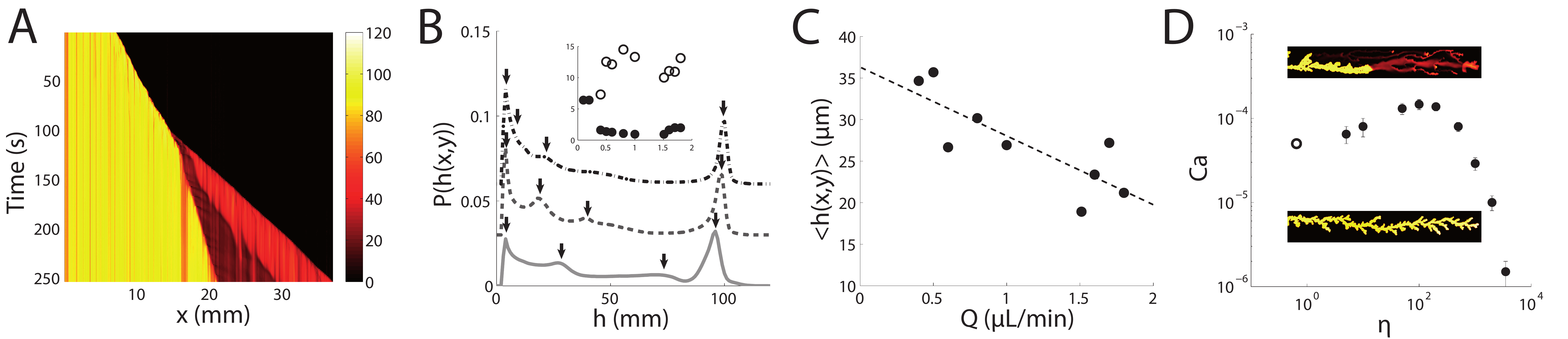}
	\caption{A-Spatiotemporal plot of $\langle h(x,t)\rangle_y$ for $\eta=10^3\,\rm cp$ and $Q=1.5\,\rm \mu l/s$. B- Probability distribution functions of the water-pattern thickness. Full line: $Q=1.8\,\mu\rm l/s$, dashed line: $Q=1.5\,\mu\rm l/s$, dotted line: $Q=0.5\,\mu\rm l/s$,  $\eta_{\rm oil}=10^3\,\rm cp$.   The three curves are shifted by a constant offset to facilitate the reading. Inset: normalized perimeter-over-area ratio  plotted as a function of $Q$ . Main finger region: filled symbols. Thin-film region: open symbols. The ratio is normalized by the one corresponding to an ideal Saffman-Taylor finger spanning one half of the channel width. C-Mean film thickness plotted versus the water flow rate. D- Phase diagram. Symbols: critical (local) capillary number at which entrainement occurs. Error bars: one standard deviation.}
\label{fig2}
\end{figure*}

The experiment is thoroughly described  in a supplementary document~\cite{supp}. Briefly, it consists in injecting a coloured aqueous solution in a  microfluidic Hele-Shaw channel  filled with silicon oil of viscosity $\eta_{\rm oil}$ ranging from 5 cp to 3500 cp. The invasion patterns are observed with a CCD camera with a spacial resolution of $12\, \rm{\mu m /pxl}$. To  gain more knowledge about their 3D morphology, we also convert  the transmitted-light intensity into the local water-pattern thickness with a resolution of $5\, \rm{\mu m}$~\cite{Saintyves:2013dc}. 
The  channels are made by bonding two glass slides with a double-sided tape. Prior to assembly,  a thin film of thiolene-based resins  is deposited on the two glass slides (NOA 81, Norland Optical Adhesives). Using  NOA81 surfaces, the advancing contact angle of the aqueous solution immersed in silicon oil  can be continuously varied from $\theta_0=120\pm2^\circ$ down to $\theta_0=7\pm2^\circ$ by means of  a  UV exposure~\cite{Levache2012}. The width and the length of the main channel are $W = 5\,\rm mm$, and $L=4.5\,\rm cm$ respectively, Fig.~1. The channel height is constant over the entire device, and is left unchanged as the fluids flow, $H=100\, \rm \mu m$.   In order to avoid any possible modification of the wetting properties, we make sure  that the main channel does not contact any aqueous liquid prior to the imbibition experiment. To do so, 
 the device is filled following a systematic sequence of injection steps described in~\cite{supp}. In addition, the chips are {\em not} recycled. More than 50 chips were used to produced the dataset introduced below. 

We first show in Fig.~\ref{fig1}A the result of a  standard drainage experiment, where silicon oil of viscosity $\eta_{\rm oil} = 10^{3}\,{\rm cp}$ is displaced in an  hydrophobic channel ($\theta_0=120^\circ$). The wedge-shape entrance of the main channel promotes  tip splitting  in this typical Saffman-Taylor  pattern~\cite{Lajeunesse2000}. The colors of Fig.~\ref{fig1}A code for the local water thickness,  the two fingers clearly fill the gap of the shallow channel. We also note that they grow along the side walls which they partly wet. The very same type of  finger shapes were observed over a decade of flow rates : $0.2\,\rm{\mu l/min}<Q<1.8\,\rm{\mu l/min}$.  
In contrast,  Fig.~\ref{fig1}B and supplementary movie 1  correspond to an imbibition experiment  ($\theta_0=7^\circ$) performed at small flow rate. The marked difference between these two fingering patterns clearly reveals the  impact of $\theta_0$ on the water-front dynamics. The   branching level is significantly increased while the width of the fingers is  reduced compared to the drainage regime.  
More surprisingly, increasing the water flow rate above $Q^\star=0.4\,\rm{\mu l/min}$,  the imbibition dynamics does not reduce to the mere propagation of a sharp water front any more,  see Figs.~1C and 1D and supplementary movies 2 and 3. Thin water films  are entrained from the finger tip throughout the oil phase,  and merge to form  complex interconnected patterns.  Increasing the flow rate,  the number of narrow thin films increases.  
Using a microscope and a 20x objective we found that the films propagate along the top and bottom walls. At this point we shall note that this latter observation is at odds with the entrainment dynamics reported in~\cite{Aarts2013}, where thin films were entrained in between the two confining walls. 

Fig. 2A conveys a clear picture of the interface dynamics at large scales. The imbibition-pattern thickness averaged over the $y$-direction, $\langle h(x,t)\rangle_y$, is plotted as a function of time and of the $x$-position along the channel. At $t=0$, the flow rate is smaller than $Q^\star$, and a branched finger grows at a constant speed. As $Q$ is increased above $Q^\star$, a thin water film is entrained and forms a rim. This rim is separated by the initial finger by an even thiner flat film. The rim moves at a constant speed  ahead the initial thick finger, which keeps on growing at a constant, yet smaller velocity.  
The  main water finger  slowly meanders  in the channel following the interconnected track left  by the entrained films thereby traping small oil pockets in the channel.
The  topology of the resulting holey imbibition pattern, Fig.~1D is not akin to the branched structure emerging from a Laplacian growth process as  observed in all the drainage experiments.  

To further characterize the pattern heterogeneities, we  measured the instantaneous distribution  ${\cal P}(h(x,y),t)$ of the film-thickness field. ${\cal P}(h,t)$ was found to be stationary, in agreement with  the  constant speed of the two fronts separating the three regions (finger, flat film and rim) in Fig. 2A. ${\cal P}(h)$ is typically composed of four peaks, that may overlap, Fig. 2B. The leftmost peak corresponds to the edges of the pattern where the water thickness is by definition vanishingly small. The second peak corresponds to the flat-film regions. The third and broadest peak is centered on the typical rim-thickness value. The rightmost narrow peak  located at $h=H$ corresponds to the main finger. The strong increase with $Q$  of the lefmost-peak amplitude  reflects  the increase of the perimeter-over-surface ratio at high water injection rate. This ratio is plotted in Fig.~2B inset both for the main finger  and for the entrained film region. As $Q$ exceeds $Q^\star$, this quantity drops discontinuously for the main finger as liquid entrainement suppresses branching. Oppositely, in the entrained-film region the ratio jumps to higher values as the  holes  increase the pattern perimeter  while reducing its  area. %
Fig.  2C also demonstrates that the mean  thickness of the films $\langle h(x,t)\rangle_{x,y}$ decreases linearly with $Q$.    

To gain more physical insight, we henceforth describe the imbibition process in term of the three dimensionless numbers that control the interface dynamics: the advancing contact angle $\theta_0$, the viscosity ratio of the two fluids $\eta\equiv\eta_{\rm oil}/\eta_{\rm water}$, and the capillary number $Ca=\eta_{\rm water} V/\gamma$, where $V$ is the interface velocity, and $\gamma$ is the surface energy of the two-fluids interface which we estimate to be $\gamma\sim 20\,\rm mPa.s$. Here we focus on the  roles of $Ca$ and $\eta$ for a small contact angle value. Repeating the same experiment with oils of different viscosity, we measured the  $z$-averaged tip velocity of the fingers from which a water film is entrained. These measurements define the experimental phase diagram shown in Fig. 2D. Unexpectedly, the critical capillary number $Ca^\star$ above which the meniscus is unstable,  undergoes non-monotonic variations with $\eta$ and displays a maximum for $\eta\sim 100$.   This observation rules out a naive scaling  argument which would consist in comparing the magnitude of the Laplace pressure and of the viscous stress in the oil (reps. in the water) phase at the macroscopic scale $H$.   Such  estimates would result in  a scalings $Ca^\star\sim 1/\eta$  (resp. $Ca^\star\sim 1$),  none of which is experimentally observed.  To go beyond this  oversimplified description, we now introduce a minimal model which accounts  for the interplay between the fluid flows and the meniscus shape at all scales. 
 \begin{figure}
	\includegraphics[width=\columnwidth]{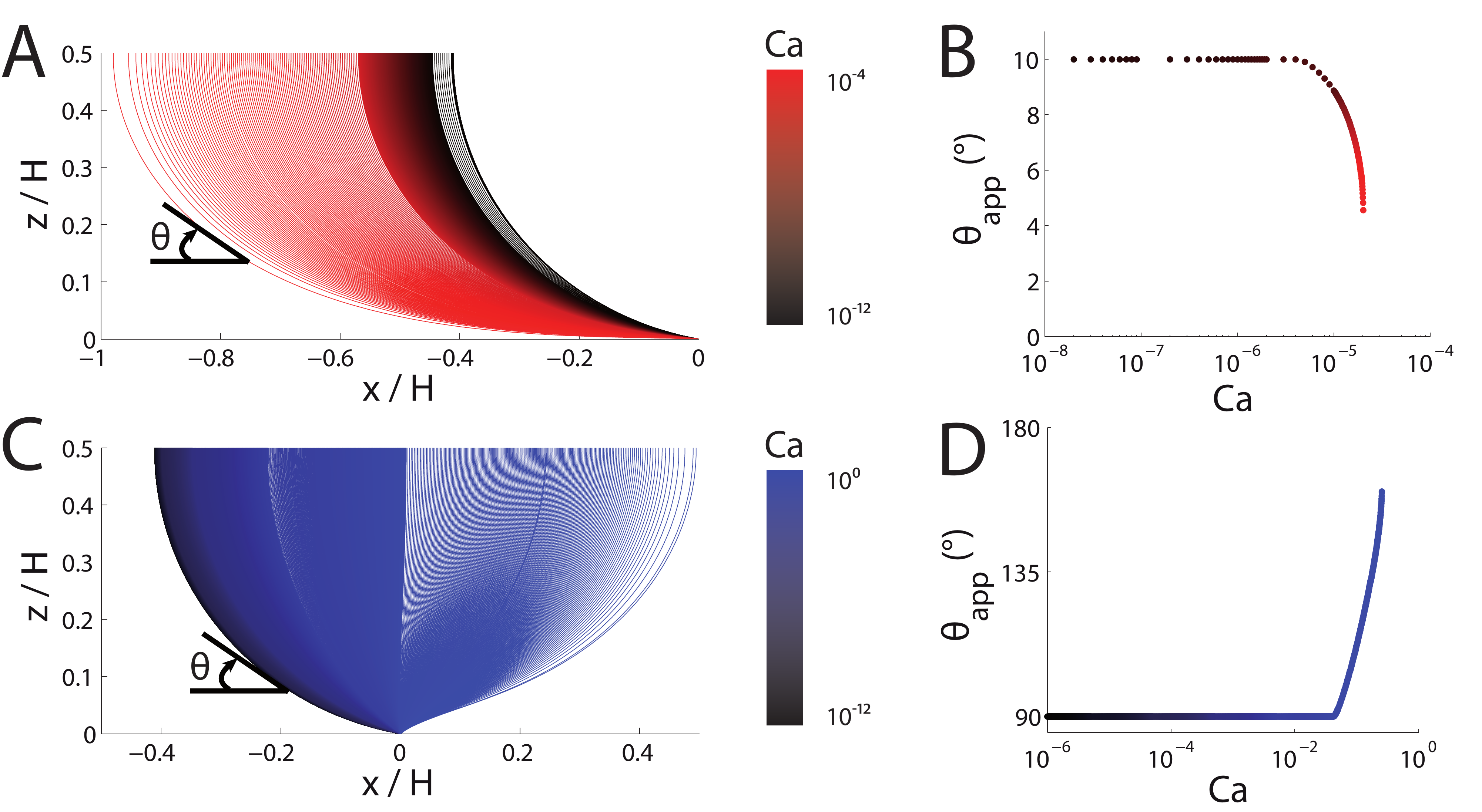}
	\caption{A-Computed meniscus profile for increasing values of $Ca$ (from black to red). $\theta_0=10^\circ$, $\eta=10^3$, $\lambda/H=10^{-5}$. B- Variations of the apparent contact angle with $Ca$. Same parameters as in A. C-Computed meniscus profile for increasing values of $Ca$ (from black to blue). $\theta_0=10^\circ$, $\eta=10^{-2}$, $\lambda/H=10^{-5}$. D-Variations of the apparent contact angle with $Ca$. Same parameters as in C}
\label{fig3}
\end{figure}

For sake of simplicity we ignore curvature effects in the $xy$-plane, and focus on steadily moving interfaces that are translationally invariant along the $y$-direction. 
 The meniscus shape is determined by the {\em local} balance between the Laplace pressure and the normal-stress discontinuity across the fluid interface. Introducing  the curvilinear coordinate along the interface, $s$, and the local interface curvature $\kappa$,  the unit-vector normal to the surface ${ \bf \hat n}$, it takes the compact form:
\begin{equation}
\gamma \kappa(s){ \bf \hat n}=\Delta  {\boldsymbol \sigma}\cdot{ \bf \hat n},
\label{eq1}
\end{equation} 
where, the $\boldsymbol \sigma$ is the  stress discontinuity  at the interface. This equation  couples to the Stokes equations for the two fluid flows.  To solve this  demanding problem, we built on \cite{Snoeijer:2006cl,Marchand2012}, and make an additional ansatz which  has proven to yield excellent agreement with  lattice Boltzman simulations~\cite{Chan:2013gc}.  In the frame moving with the contact line, the velocity and the pressure fields in both phases are assumed to be locally given by the Hu and Scriven solutions for the flow in a wedge of  angle $\theta(s)$,  where $\theta(s)$ is the local angle between the tangent vector and the $x$-axis, Fig.~3A~\cite{Huh1971}. Within this approximation, the stress discontinuity in Eq.~\ref{eq1} is readily computed, and the shape of the interface is fully prescribed by completing Eq.1 with the boundary conditions: $\theta(s=0)=\theta_{0}$, $\theta(s=\ell/2)=\pi/2$, where $\ell$ is the curvilinear length of the meniscus.  
 Eq.~\ref{eq1} is then recast  into a 4-dimensional dynamical system, and this boundary value problem is   effectively solved using an iterative collocation method as explained in~\cite{supp}.

The evolution of the meniscus shape with the capillary number is shown in Fig.~3A  for  $\eta=10^3$. Increasing $Ca$  increases the meniscus length and   reduces of the apparent contact angle value. More quantitatively, we show in Fig.~3B that $\theta_{\rm app}$,  measured here at the  point of minimal curvature,  decays to $0$ for a finite value of $Ca$ above which no stationary solution is found for the meniscus profile:  in  agreement with our experimental findings, a low-viscosity-liquid film is entrained along the walls. 

However, when considering the  case of  moderate and small viscosity ratios, we found another instability mechanism, see e.g. Fig. 3C for which $\eta=10^{-2}$. As $Ca$ increases, the apparent curvature of the meniscus decreases and changes sign. As a result the apparent contact angle increases toward $\pi$. Above a critical $Ca$ value, again, no stationary solution is found. However, this dual instability yields a meniscus shape opposite to the one found for large $\eta$: a liquid sheet grows upstream between the two plates. The interface profile shown in Fig.~3C exactly corresponds to the one one reported in~\cite{Aarts2013} for colloidal liquid with moderate viscosity contrast, $\eta=2.7$. Therefore our numerical results solve the apparent contradiction between~\cite{Aarts2013} and our experimental findings:  viscous-finger menisci can experience two qualitatively different liquid-entrainment instabilities.

To further check the consistency of our  predictions.  We conducted experiments with   silicon oil having an ultra-low viscosity $\eta=0.65$. Even though this viscosity ratio prevents the formation of viscous fingers, we did observe a strong change in the liquid motion at sufficiently high $Ca$. Again, above a  critical capillary number (open symbol in Fig. 2D), a liquid sheet is entrained between the two plates ahead the main front, and subsequently re-wets the confining wall. As a result  oil droplets are trapped on the two solid surfaces, see  supplementary movie 4. This observation is akin to the ones reported both in~\cite{Aarts2013}, and in~\cite{Marchand2012} for air entrainment in a liquid bath. Together with our first experimental findings,  this  last experiment unambiguously confirms that,  thin films can be entrained from a driven meniscus according to two different scenarios set by the magnitude of the viscosity ratio, . 

We stress that both scenarios  echoes the intricate coupling between the two fluid flows at the contact line. Even when it  is associated to the smaller viscosity, the flow in the  wetting phase significantly alter the stability of the meniscus upon imbibition dynamics. Sufficiently close to the contact line, due to the geometrical divergence of the strain rate,   $\sigma_{\rm water}$  compares to  $\sigma_{\rm oil}$. In the absence of any intrinsic length scale for the interface dynamics in Eq. 1, and in the Stokes equation, the local modification of the meniscus curvature by the flows at the tip of the liquid wedge  propagates up to the macroscopic scales. 
 \begin{figure}
	\includegraphics[width=\columnwidth]{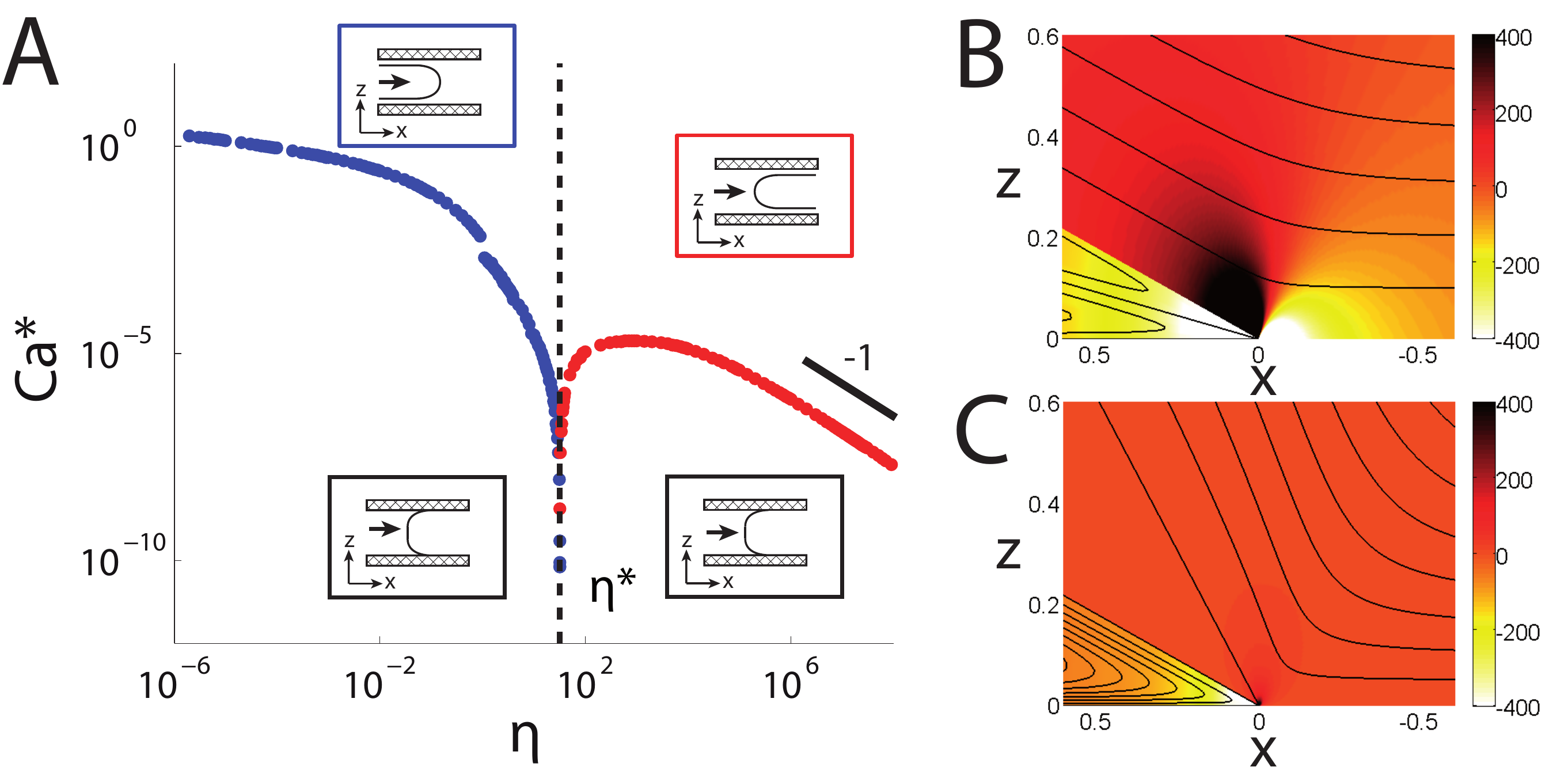}
	\caption{A-Stability diagram for a liquid interface driven in a Hele-Shaw channel: numerics. $\theta_0=10^\circ$, $\lambda/H=10^{-5}$. Insets: sketches of the interface shape. B-Stream lines and pressure field in a wedge of angle $\theta_0=20^\circ$ for $\eta/\eta^\star=10$. Note the existence of a depression ahead the contact line. C-Stream lines and pressure field in a wedge of angle $\theta_0=20^\circ$ for $\eta/\eta^\star=0.1$. The color codes for the local pressure. Arbitrary unit.}
\label{fig4}
\end{figure}

These  two entrainment scenarios define  the stability diagram plotted in Fig.~4A. The stable meniscus region in the $(\eta,Ca)$ plane is bounded by  two critical curves, that meets at $\eta=\eta^\star$. Below $\eta^\star$ the entrained films propagate at the center of the gap, whereas above $\eta^\star$  entrainement occurs along the confining walls.   This prediction capture well the salient features of the experimental phase diagram shown in Fig.~2D. However we did not achieve a quantitative agreement. For instance $\eta^\star$ was predicted to be of the order of $100$, yet it was measured to be close to unity. Needless to say that this discrepancy is not really surprising given the  simplification of the meniscus geometry in the $y$-direction, and potentially due to pining effect at the contact line which we have ignored. 

Two last comments are in order. Firstly, we provide a simple criteria  to distinguish between the two liquid-entrainment scenarios. To do so, we consider the flow in a perfect wedge of angle $\theta_0$, which is a reasonable approximation in the very vicinity of the contact line.  Below $\eta^\star$, in the low-viscosity liquid, the stream lines have a simple V-shape, Fig. 4C. Conversely above $\eta^\star$ they split into  two recirculations, Fig. 4B. Simultaneously the radial velocity of the fluids at the interface changes its sign: $\eta^\star$ is  defined as the viscosity-contrast value at which the radial component of the interface velocity vanishes. Secondly, looking now at the pressure field in the oil phase, we can gain additional physical insight into the high $\eta>\eta^\star$ regime.  Fig. 4B indeed reveals that  the tip of the liquid wedge is pulled downstream by a marked depression spot  located at $z=0$ in the oil phase, thereby promoting entrainment past the solid wall.

In summary, we have demonstrated  a novel class of forced imbibition patterns. They  stem from the entrainment of thin films out of the interface between a wetting fluid and a high viscosity fluid when driven past solid surfaces. In addition, we have introduced a minimal theoretical framework which accounts well for all the  imbibition-induced meniscus instabilities that have been reported so far. Our findings should provide useful guidelines for the formulation of effective additive for cleaning, and enhance oil recovery applications.

We thank B. Andreotti, J. Snoeijer and E. Santanach Carreras for illuminating discussions, and C. Odier for help with the experiments. D.B acknowledge support from Institut Universitaire de France. 

\newpage
{\centerline {\bf SUPPLEMENTARY INFORMATIONS}}
{\section{Experiments}

\subsection{Microfluidics}
\begin{figure}[h]
	\includegraphics[width=0.8\columnwidth]{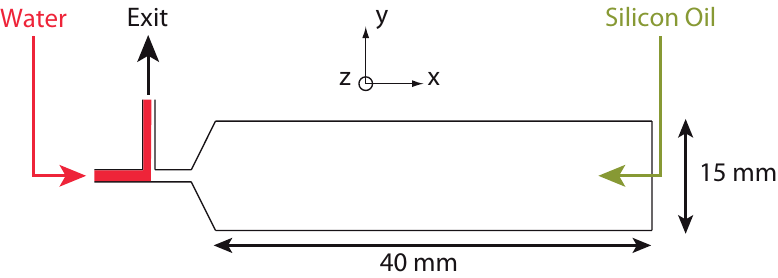}
	\caption{Sktech of the microfluidic channels, and of the initial shape of the water-silicon oil interface.}
\end{figure}
The experiment consists in injecting an aqueous solution (water, SDS 1 wt\% and food dye) in a  microfluidic Hele-Shaw channel  filled with silicon oil (Rhodorsil Oils of viscosity ranging from 5 cp to 3500 cp). 
The  channels are made by bonding two glass slides with a double-sided tape cut with a precision plotting cutter (graphtec robo). Prior to assembly,  a thin film of thiolene-based resins  is deposited on the two glass slides (NOA 81, Norland Optical Adhesives). Using  NOA81 surfaces, the advancing contact angle of the aqueous solution immersed in silicon oil  can be continuously varied from $\theta_0=120\pm2^\circ$ downto $\theta_0=7\pm2^\circ$ by means of  a deep-UV exposure~\cite{Levache2012}. 
The geometry of the resulting channels is sketched in the supplementary figure 1. Their height is constant over the entire device, $H=100\, \rm \mu m$, and does not change as the fluids are flown. The width and the length of the main channel are $W = 5\,\rm mm$, and $L=4.5\,\rm cm$ respectively. In order to  accurately control   both the wetting properties of the walls, and  the initial shape of the water-oil meniscus, the liquids are injected as follows. First, the channel is filled with silicon oil  by applying a constant $200\, \rm{mbars}$ pressure which is maintained over the entire experiment. Then, the aqueous solution is flown at a constant flow rate  with a precision seringe pump (Nemesys, Cetoni). The two fluids meet at the T-junction and form a flat interface, supplementary Figure 1. Once the  interface reaches a stationary shape,  the T-junction outlet is closed to trigger the invasion the Hele-Shaw cell by the aqueous solution.  In order to avoid any possible modification of the wetting properties  of the hydrophilic  surfaces  by the water, the surfactant, or the dye molecules the chips are {\em not} recycled. 
\begin{figure}
	\includegraphics[width=0.5\columnwidth]{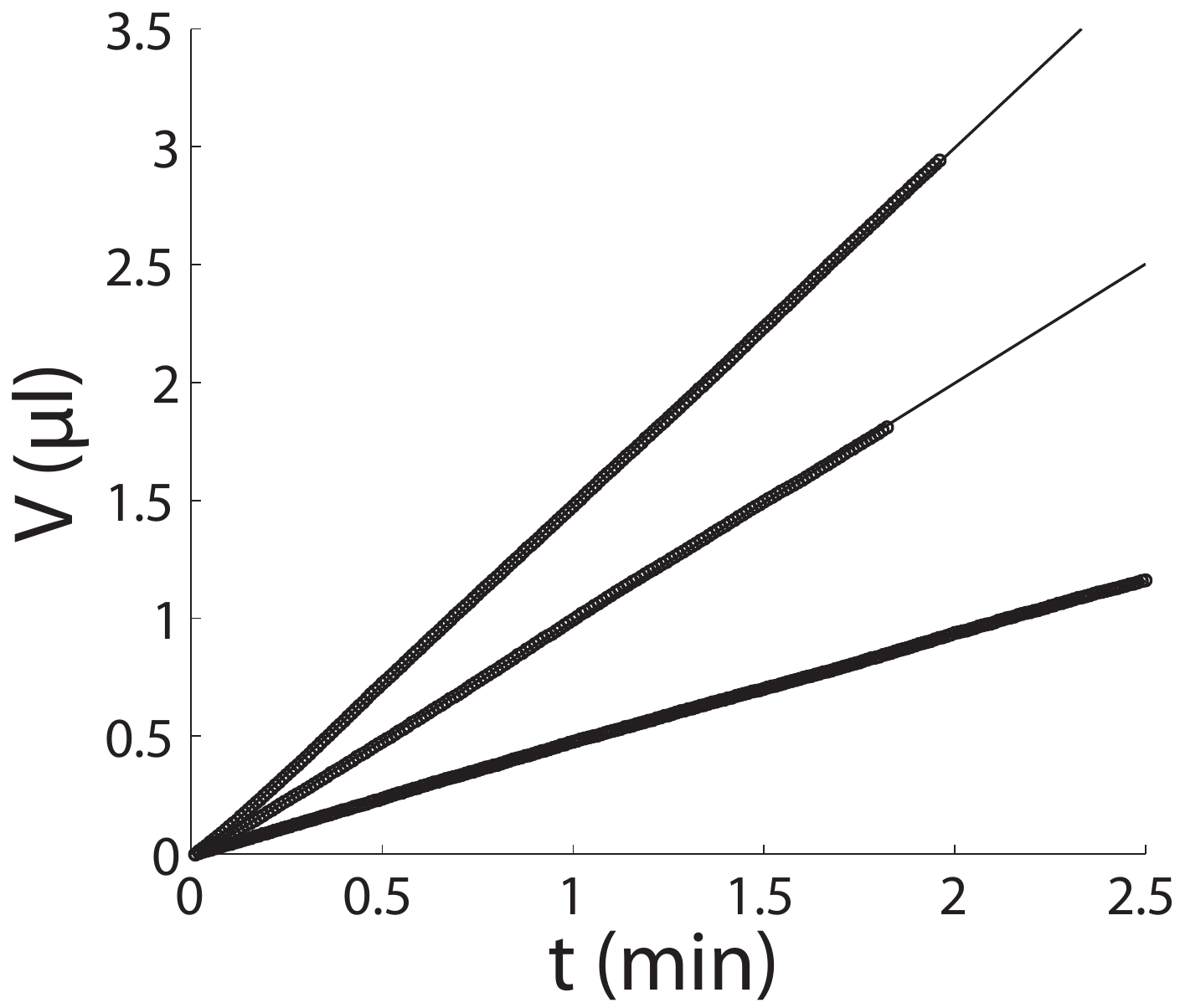}
	\caption{Circles: Measurement volume plotted as a function of  time for three imbibition experiments performed at imposed flow rate: $0.5 \, \mathrm{\mu L/min}$, $1.0 \, \mathrm{\mu L/min}$, $1.5 \, \mathrm{\mu L/min}$. Full lines: best linear fits.}
\end{figure}
\subsection{Observations and measurements}
The invasion patterns are observed with a $60\, \rm{mm}$ macro lens (Nikkor f/2.8G, Nikon) mounted on a $8\, \rm{Mpxls}$, 14bit CCD camera  (Prosilica GX3300), which yield  a spacial resolution of $12\, \rm{\mu m /pxl}$. We also convert  the transmitted-light intensity into the local water-pattern thickness with a resolution of $5\, \rm{\mu m}$. 

The local thickness $h(x,y,t)$ of the water films relate to the image intensity $I(x,y,t)$ via the Beer-Lambert absorption law:
\begin{equation}
h(x,y,t)=-\epsilon c  \ln{\frac{I(x,y,t)}{I_0(x,y,t)}}
\end{equation}
where  $c_0$is the dye concentration, $\epsilon$ is the absorptivity, and $I_0$ is the transmitted intensity for a channel filled with silicon oil.  $\epsilon$ was determined by performing experiments with colored water only, in  channels of known height using  solutions with increasing concentrations. Note that the measure of $I_0(x,y)$ allowed us to correct the spatial  heterogeneities of the observation setup. We also performed a systematic correction of the (minute) temporal fluctuations of the light source by bringing the average light intensity outside the channel to the same value for all images. We benchmarked this method by measuring the increase of the volume of aqueous solution injected at constant flow rates e.g. for the three imbibition experiments corresponding to Fig. 1 ($Q = 0.5 \,$,  $1.0$ and $1.5 \, \mathrm{\mu L/min}$respectively). These three curves shown in supplementary figure 2 are perfectly fitted by straight lines, the slope of which indeed corresponds to the imposed flow rates within a $3\%$ error (best linear fits: $Q_{\rm measured} = 0.486 \,$, $1.01$ and $1.517 \, \mathrm{\mu L/min}$ respectively).

\section{Numerical resolution of the meniscus shape}
The geometry of the meniscus is depicted in the supplementary figure 3. In order to solve Eq.~1 within a small curvature approximation and a wedge-flow ansartz~\cite{Chan:2013gc}, we first consider $\ell$ as an extra variable, and rescale $s$ by $\ell$, and the local thickness $h(s)$ by $H$, the interface shape is then the solution of:
\begin{align}
	\dot h(s) &= \ell \sin{\theta}, \\
	\dot \theta (s) &= \ell\, \kappa, \\
	\dot  \kappa(s) &= \frac{3\ell Ca  }{h \left ( h + 3 \lambda_l \right ) } f \left ( \theta , \eta \right ), \label{f} \\
	\dot \ell(s)  &= 0,
\end{align}
associated with  the boundary conditions:  $h(0)=0$, $h(1)=1/2$, $\theta(0)=0$, and $\theta(1)=\pi/2$. $\lambda_l\ll H$ is a slip length used to regularize Eq.~4, none of the following results qualitatively depends on this parameter. The non-linear function $f(\theta,M)$ in Eq.~\ref{f} is readily computed from the Hue and Scriven solution for a wedge flow~\cite{Huh1971}:
\begin{align}
f \left ( \theta , R \right ) &\equiv \frac{ 2 \sin^3{\theta} \left [ R^2 f_1 \left ( \theta \right ) + 2 R f_3 \left ( \theta \right ) + f_1 \left ( \pi - \theta \right ) \right ]}{3 \left [ R f_1 \left ( \theta \right ) f_2 \left ( \pi - \theta \right ) - f_1 \left ( \pi - \theta \right ) f_2 \left ( \theta \right ) \right ] }\nonumber \\
f_1 \left ( \theta \right ) &\equiv \theta^2 - \sin^2{\theta} \nonumber \\
f_2 \left ( \theta \right ) &\equiv \theta - \sin{\theta}\cos{\theta} \nonumber \\
f_3 \left ( \theta \right ) &\equiv \theta \left ( \pi - \theta \right ) + \sin^2{\theta}.
\end{align}
This boundary value problem is  effectively solved using an iterative collocation method. In fact we used the bvpc4 Matlab routine~\cite{shampine2003solving}. 
\begin{figure}[b]
	\includegraphics[width=0.5\columnwidth]{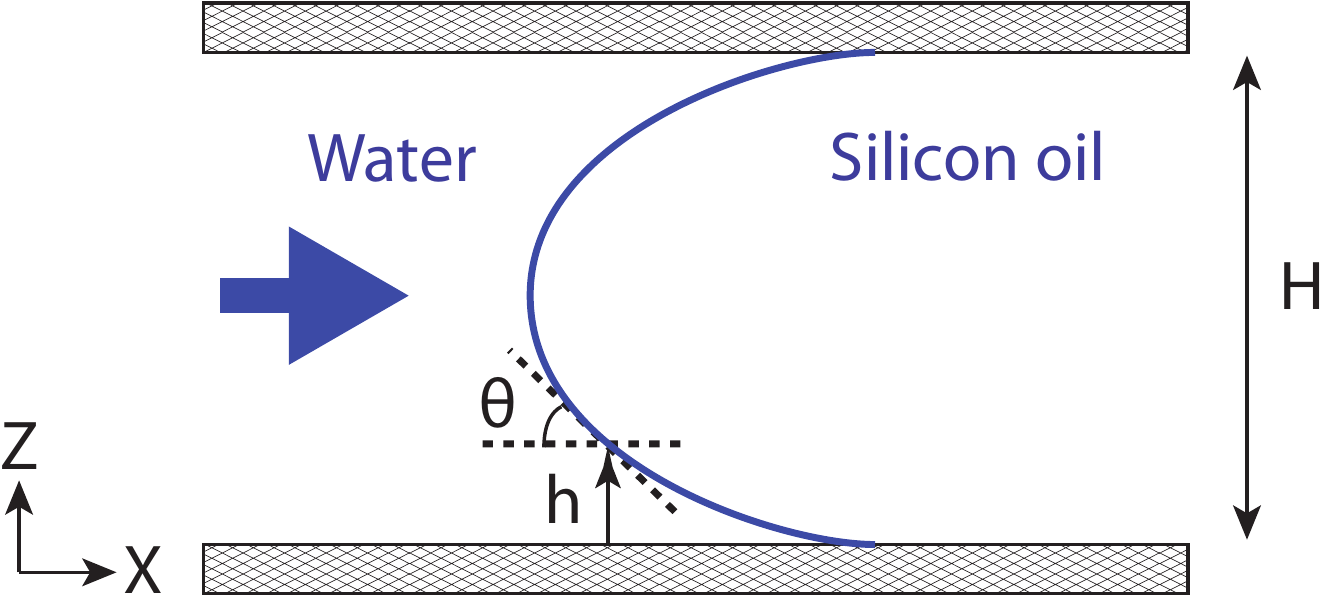}
	\caption{Sktech of the microfluidic channels, and of the initial shape of the water-silicon oil interface.}
\end{figure}

\end{document}